\documentclass[aps,pra,reprint]{revtex4-1}
\usepackage{lipsum}
\usepackage{graphicx}
\usepackage{epstopdf}
\usepackage{dcolumn}
\usepackage{bm}
\usepackage{physics}
\begin{document}
\preprint{APS/123-QED}

\title{Generation and distribution of atomic entanglement in coupled-cavity arrays}
\author{J. P. Mendon{\c{c}}a}
\email{jpedromend@gmail.com}
\affiliation{%
 Instituto de F\'{i}sica, Universidade Federal de Alagoas, 57072-900 Macei\'{o}, AL, Brazil
}
\author{F. A. B. F. de Moura}
\affiliation{%
 Instituto de F\'{i}sica, Universidade Federal de Alagoas, 57072-900 Macei\'{o}, AL, Brazil
}
\author{M. L. Lyra}
\affiliation{%
 Instituto de F\'{i}sica, Universidade Federal de Alagoas, 57072-900 Macei\'{o}, AL, Brazil
}
\author{G. M. A. Almeida}
\affiliation{%
 Instituto de F\'{i}sica, Universidade Federal de Alagoas, 57072-900 Macei\'{o}, AL, Brazil
}

\date{\today}

\begin{abstract}
We study the dynamics of entanglement in a 1D
coupled-cavity array, each cavity containing a two-level
atom, via the Jaynes-Cummings-Hubbard (JCH) Hamiltonian in the single-excitation sector. 
The model features a rich variety of 
dynamical regimes that can be harnessed for entanglement control. 
The protocol is based on setting an
excited atom above the ground state and further letting it evolve
following the natural dynamics of the Hamiltonian. 
Here we focus on the
concurrence between pairs of atoms and its relation to
atom-field correlations and the structure of the array. 
We show that the extension and distribution pattern of 
pairwise entanglement can be manipulated
through a judicious tuning of the atom-cavity coupling strength only. 
Our work offers a comprehensive account over the  
machinery of the single-excitation JCH Hamiltonian as well as
contributes to the design of 
hybrid light-matter quantum networks. 
\end{abstract}

\maketitle

\section{\label{sec1}Introduction}

Quantum entanglement is one of the most intriguing properties 
of nature with no classical analog \cite{horodecki2009}. 
It is a key manifestation in many-body physics for it plays
a significant role in quantum phase transitions \cite{amico2008,vidal2003,osterloh2002}.
In addition, entanglement is a fundamental resource in quantum information 
processing tasks such as teleportation \cite{bennett1993}, quantum cryptography 
\cite{ekert1991,gisin2002} and quantum dense coding \cite{bennett1992}, to name 
a few. 
In this respect, in order to properly design such a class of protocols one must 
be able to faithfully transmit quantum states and establish entanglement over 
arbitrarily distant parties (qubits) \cite{divincenzo1995,cirac1997}.
Setting reliable quantum communication channels is thus a primary step  
towards building large-scale quantum networks \cite{kimble2008, schoelkopf2008}.

%
Along those lines, 
photonic channels stand out as current technology allows for light propagation
over large distances with negligible decoherence. 
On top of that, local quantum information processing units (nodes) may
consist of single atoms placed in optical resonators. This allows for
light-matter
interfacing with high degree of control, thanks to experimental advances
in cavity-QED-based architectures \cite{ritter2012, nolleke2013, reiserer2014, reiserer2015}.

A paradigmatic framework to deal with coupled-cavity systems 
is the Jaynes-Cummings-Hubbard (JCH) model, where cavities containing
single two-level atoms
are brought 
together enough to allow for photon tunneling. Atom-cavity
coupling is given by the acclaimed Jaynes-Cummings interaction in the rotating-wave approximation.  
Early developments of the model initiated with the discovery that it displays a 
superfluid to Mott insulator quantum phase transition \cite{greentree2006, angelakis2007}.
This established coupled-cavity systems also as potential many-body quantum simulators \cite{tomadin2010}. Furthermore, 
the hybrid light-matter nature of the excitations unveils novel phases of matter \cite{rossini07} and can also be useful
for quantum information processing tasks \cite{almeida2016} (for reviews on the model and related content, see Refs. \cite{hartmann2008,tomadin2010}).   

%

In this work, we further explore the versatility of a 1D coupled-cavity array in order to 
generate and distribute entanglement, which is a key element in the design of quantum networks \cite{kimble2008}. 
The protocol is based on 
preparing a impurity -- here, an excited atom -- over a well-defined ground state and let it evolve following the natural,
Hamiltonian dynamics of the system. Along the process it is possible to generate entanglement as shown for spin chains in Refs. \cite{amico2004, almeida2017} (cf. \cite{fukuhara2015} for an experimental realization).   
Here the initial atomic excitation is released from the middle of a coupled-cavity array 
prepared in the vacuum state (no photons) with all the remaining atoms in
their ground state.
As the JCH model commutes with the total number operator, the dynamics ends up being
restricted to the single-excitation subspace what allows for easy analytical treatment \cite{makin2009} in addition to displaying very rich properties \cite{makin2009, ciccarello2011, almeida2016, almeida2017}.

We carry out a detailed analysis over limiting interaction regimes of the JCH Hamiltonian
and track entanglement evolution over time in two forms: 
the von Neumann entropy for the whole atomic component in regard to 
the photonic degrees of freedom and the concurrence between atomic pairs.  
We discuss the role of atom-field entropy 
in establishing atomic entanglement and how its spatial distribution 
profile is related to the atom-cavity interaction strength and
the structure of the embedded array.

In the following Sec. \ref{sec2}, we introduce the JCH Hamiltonian. In Sec. \ref{sec3}, the weak- and strong-coupling regimes
of the model is addressed in detail. In Sec. \ref{sec4} we outline the entanglement signatures of interest and
discuss their dynamics focusing on those limiting regimes. Conclusions are drawn in Sec. \ref{sec5}.

\section{\label{sec2}Jaynes-Cumming-Hubbard model}

We consider an one-dimensional array of $N$ high-quality coupled cavities, each containing a single two-level atom, with $\ket{g}$ and $\ket{e}$
denoting the ground and excited states, respectively. Each atom 
interacts with the field through the local Jaynes-Cummings (JC) Hamiltonian (in the rotating-wave approximation) \cite{jaynes1963}
\begin{equation}\label{jc}
H_{x}^{\mathrm{JC}} = \omega_{c} a_{x}^{\dagger}a_{x} + \omega_{a} \sigma_{x}^{+}\sigma_{x}^{-} +
g(\sigma_{x}^{+}a_{x}+\sigma_{x}^{-} a_{x}^{\dagger}),
\end{equation}
where
$ a_{x}^{\dagger} $ ($a_{x}$) and $\sigma_{x}^{+}$ ($ \sigma_{x}^{-} $)
are, respectively, the bosonic and atomic raising (lowering)
operators acting on the $x$-th cavity, $g$ is the atom-field coupling strength, 
$\omega_{c}$ is the cavity frequency, and
$\omega_{a}$ is the atomic transition frequency. We set $\hbar = 1$ for convenience.
The eigenstates of Hamiltonian (\ref{jc}) are dressed (hybrid) states featuring photonic
and atomic excitations known as polaritons, which in resonance ($\Delta = \omega_{a}-\omega_{c}=0$) read $\ket{n\pm}_{x} = (\ket{g,n}_{x}\pm\ket{e,n-1}_{x})/\sqrt{2}$ with energies $E_{n}^{\pm} = n\omega_{c}\pm g\sqrt{n}$, where $\ket{n}_{x}$ denotes a $n$-photon
Fock state at the $x$-th cavity. Note that the vacuum state $\ket{g,0}_{x}$ is also an eigenstate, 
with zero energy.

We now assume that the local cavity modes overlap
in such a way allowing photonic tunnelling in an uniform array.
This coupled-cavity system is described by the JCH Hamiltonian
\begin{equation} \label{jchhamilt}
H=\sum_{x=1}^{N}H_{x}^{\mathrm{JC}} - J\sum_{x=1}^{N-1}(a_{x+1}^{\dagger}a_{x}+\mathrm{H.c.}),
\end{equation}
with $J$ being the photon tunnelling. 
Hereafter we fix $\omega_{c} = 0$, which is equivalent to adjust the whole
free-field normal-mode spectrum around zero (thus, the detuning is set by $\omega_{a}$ only).
The above Hamiltonian acts on basis states of the form
$\bigotimes_{x=1}^{N}\ket{s,n}_{x}$, with $s\in \lbrace g,e \rbrace $. Sorting out these states
according to the total excitation number, Hamiltonian (\ref{jchhamilt}) can be 
expressed by $H=\mathrm{diag}[H^{(0)},H^{(1)},H^{(2)},\ldots]$, where $H^{(j)}$ denotes
the Hamiltonian matrix spanned on basis states 
featuring a fixed number $j$ of excitations.

Here we focus on the generation of entanglement out of 
localized atomic state $\ket{e}_{i}$
with all the remaining atoms in their ground state and no photons.
In both cases, the system dynamics is restricted to the single-excitation subspace, $H^{(1)}$, which is spanned by
$|1_x\rangle\equiv \hat a_{x}^{\dagger}\left| \emptyset \right>$ 
and $|e_x\rangle \equiv \hat \sigma_{x}^{+}\left| \emptyset \right>$,
with $\left| \emptyset \right> \equiv \ket{\mathrm{vac}}\ket{g}_{1}\cdots\ket{g}_{N}$,
where 
the former denotes a single photon at the $x$-th cavity and the latter represents the $x$-th atom excited.
The Hilbert-space dimension is thus twice the number of cavities. 

In general, Hamiltonian (\ref{jchhamilt}) 
yields rich dynamics even in the single-excitation sector \cite{makin2009, almeida2013, almeida2016}.
In the following we address two limiting regimes of interest
that will help us out to visualize the entanglement dynamics afterwards. 

\section{\label{sec3}Interaction regimes in the single-excitation sector}

%
\begin{figure}[t] 
\includegraphics[width=0.5\textwidth]{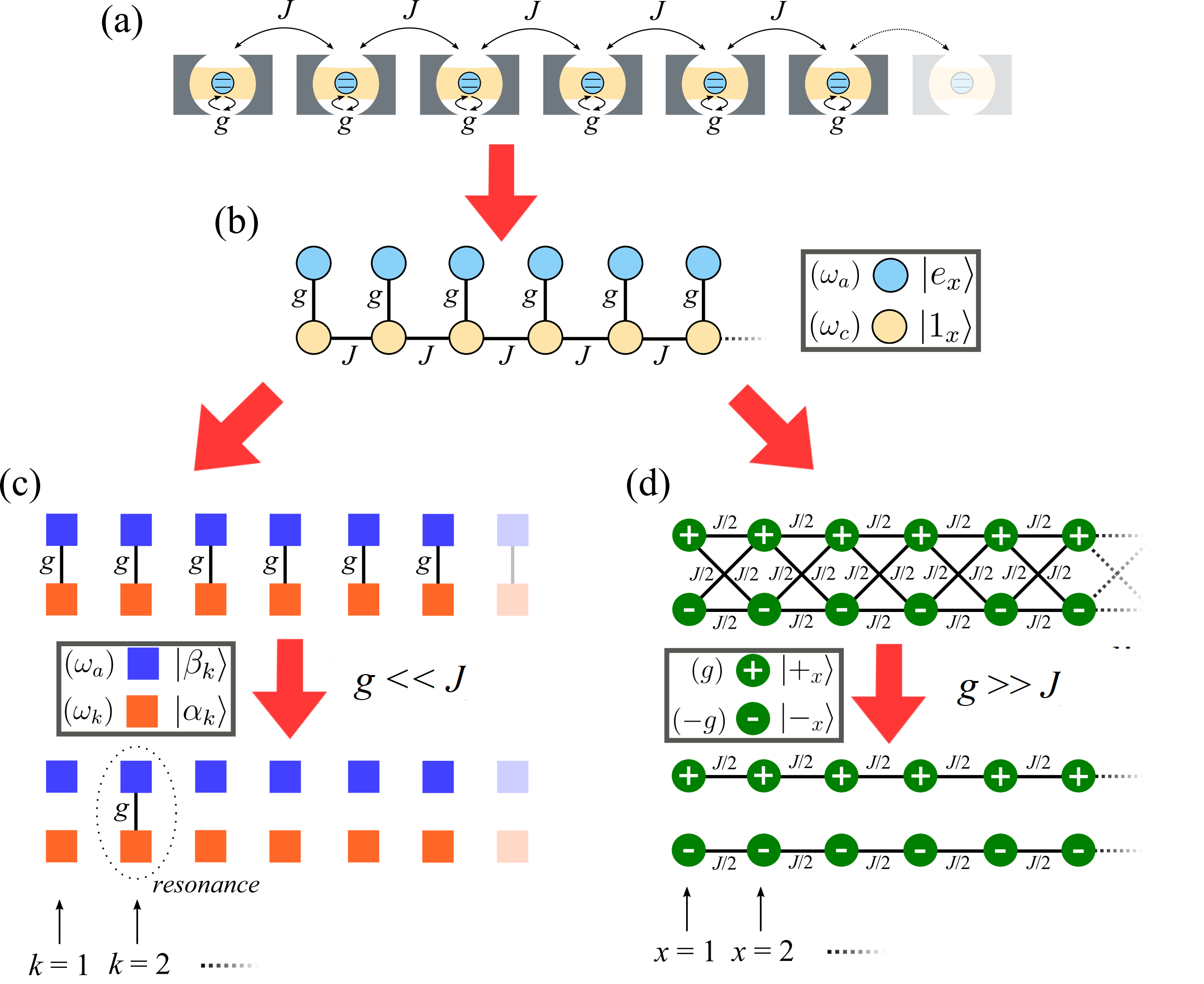}
\caption{\label{fig1} (Color online) State-graph structure of the JCH Hamiltonian describing a coupled-cavity system.
(a) Sketch of an array featuring a uniform pattern of hopping rates $J$. Each cavity has frequency $\omega_c$ and is 
coupled to a two-level atom (or qubit) with frequency $\omega_a$ through a local atom-field interaction given by $g$. 
(b) In the single-excitation sector, the model is reduced to a tight-binding chain (with extra, vertically attached sites) where single-photon states ($\{\ket{1_x}\}$) 
spreads over the array and is eventually converted into atomic degrees of freedom ($\{\ket{e_x}\}$). 
(c) The JCH Hamiltonian may be expressed in terms of 
decoupled JC-like interactions between free-field normal modes $\{\ket{\alpha_k}\}$ and its atomic 
analog $\{\ket{\beta_k}\}$. In the weak atom-field coupling regime ($g \ll J$), proper tuning leads to
a single resonant effective interaction (for $k=2$ in the example).
(d) Another form of expressing the JCH Hamiltonian is in terms of polaritonic states $\{\ket{+_x},\ket{-_x}\}$ [see Eq. \ref{Hpol}]. 
The resulting double inter-connected array can be decoupled in the strong atom-field 
coupling regime ($g\gg J$) by dropping out fast rotating terms. In this particular case, we have a standard hopping-like Hamiltonian 
for each branch of polariton having exactly the same dispersion law of the embedded structure with the hopping intensity rescaled to $J/2$. }
\end{figure}

First, we recall that when the atom-field interaction strength $g$ and the atomic transition frequency $\omega_{a}$ are uniform across the array, Hamiltonian (\ref{jchhamilt}) can be rearranged as a sum of $N$ decoupled JC-like interactions, $H = \sum_{k} H_{k}$, in terms of normal modes, where \cite{ogden2008, makin2009, ciccarello2011, almeida2016}
\begin{equation} \label{Hnormal}
H_{k} =  \omega_{k} \alpha_{k}^{\dagger} \alpha_{k}+\omega_{a} \beta_{k}^{\dagger} \beta_{k}
+ g(\alpha_{k}^{\dagger} \beta_{k} + \beta_{k}^{\dagger} \alpha_{k}),
\end{equation}
and
$\alpha_{k}^{\dagger} \equiv \vert \alpha_{k} \rangle \langle \emptyset \vert$ 
($\beta_{k}^{\dagger} \equiv \vert \beta_{k} \rangle \langle \emptyset \vert$) 
is the field (atomic) normal-mode operator.
In other words, $\lbrace \ket{\alpha_{k}} \rbrace$ is the set of $N$ eigenstates of the hopping (free-field) Hamiltonian, with eigenvalues $\lbrace \omega_{k} \rbrace$, 
each having the form $\ket{\alpha_{k}} = \sum_{x}v_{k,x}\ket{1_{x}}$.   
The atomic states $\vert \beta_{k} \rangle$ 
are set with the very same spatial profile (amplitudes) as
their photonic counterpart, that is $\ket{\beta_{k}}=\sum_{x}v_{k,x}\ket{e_{x}}$ but all 
lying at the same frequency $\omega_{a}$. Although we are dealing with a uniform pattern of hopping rates, the above situation is valid regardless of the embedded adjacency matrix.
Therefore, the model can be solved analytically once one knows the whole free-field
spectral decomposition. 
Indeed,
since the above Hamiltonian is a $2\times 2$ block-diagonal matrix indexed by $k$, its eigenstates are found to be \cite{ciccarello2011, almeida2016}
\begin{equation}\label{dressed}
\left| \psi_{k}^{\pm} \right> = A_{k}^{\pm} \left| \alpha_{k} \right>
+ B_{k}^{\pm} \left| \beta_{k} \right>,
\end{equation}
where
\begin{equation}\label{dressed_a}
A_{k}^{\pm} = \dfrac{2g}{\sqrt{(\Delta_{k} \pm \Omega_{k})^2+4g^{2}}},\,\,\,B_{k}^{\pm} = \dfrac{\Delta_{k} \pm \Omega_{k}}{\sqrt{(\Delta_{k} \pm \Omega_{k})^2+4g^{2}}},
\end{equation}
$\Delta_{k} = \omega_{a} - \omega_{k}$ is the detuning between the atomic and the field normal-mode frequency, and
$\Omega_{k} = \sqrt{\Delta_{k}^{2} +4g^{2}}$ is the corresponding vacuum Rabi frequency. 
The energy levels are given by
\begin{equation}\label{dressed_energy}
\varepsilon_{k}^{(\pm)}=\tfrac{1}{2}\,(\omega_a+\omega_{k} \pm \Omega_{k}).
\end{equation}

The JCH Hamiltonian written in the form of effective JC interactions [see Eq. (\ref{Hnormal})] allows for a convenient visualization of the system's behavior, as shown in Fig. \ref{fig1}(c). One of
the most interesting features is the possibility of setting up a particular mode to 
trigger out a pair of dressed JC-like states [c.f. Eq. (\ref{dressed})].
This can be done in the weak atom-field coupling regime, $ g \ll J$, upon a judicious tuning of the 
atomic frequency $\omega_{a}$. In order to see this, let us move on to the interaction picture, 
\begin{equation}
H^{I}(t) = g\left( \sum_{k} \alpha_{k}^{\dagger}\beta_{k}e^{-i\Delta_{k} t} + \mathrm{H.c.}\right).
\end{equation}
Setting $\omega_{a}$ in resonance with a given mode, say $k'$, one of the terms becomes
time-independent ($\Delta_{k'} = 0$) and 
considering $g\ll\lbrace\Delta_{k\neq k'}\rbrace$, all the remaining terms become fast rotating and thus can be ignored. Going back to the Schr\"{o}dinger picture,  
we are left with the effective Hamiltonian
\begin{equation} \label{Heff1}
H_{\mathrm{eff}} = H_{k'} + \sum_{k\neq k'} (\omega_{k} \alpha_{k}^{\dagger} 
\alpha_{k}+\omega_{a} \beta_{k}^{\dagger} \beta_{k}),
\end{equation}
where the first term is given by Eq. (\ref{Hnormal}). The above equation
describes a single JC-like interaction taking place at
mode $k'$, spanning a pair of fully dressed states $\ket{\psi_{k'}^{\pm}}$ 
[cf. Eq. (\ref{dressed})], 
with all the other atomic and field modes uncoupled. 
A schematic representation of this regime is 
shown in Fig. \ref{fig1}(c). 

Within the weak coupling regime, if an atomic excitation is prepared 
somewhere along the array, say $\ket{\psi(0)} = \ket{e_{x_{0}}}$ , it may get stuck 
depending on the nature of the free-field spectrum and resonance conditions \cite{makin2009, ciccarello2011, almeida2013,almeida2016}. In the off-resonant case, that is $\omega_{a} \neq \omega_{k}$ for all $k$, it is immediate to note that $\ket{\psi(t)} = e^{-iHt}\ket{\psi(0)} = e^{-i\omega_{a}t}\ket{e_{x_{0}}}$ and so the atomic excitation indeed freezes at the initial cavity $x_{0}$.
Now, if $\omega_a$ is put in narrow resonance with 
a given (nondegenerate) mode $k'$
thereby setting up 
a JC-like interaction between
this mode and its atomic counterpart, 
the evolved atomic coefficients read
\begin{equation}\label{at_amp}
c_{a,x}(t) = e^{-i\omega_{a}t}\left[\sum_{k\neq k'}v_{k,x}v_{k,x_{0}}^{*}+\cos{(gt)v_{k',x}v_{k',x_{0}}^{*}}\right],
\end{equation}
and as $\sum_{k} v_{k,x}v_{k,x_{0}}^{*} = 1$ ($=0$) for $x= x_{0}$ ($x \neq x_{0}$) 
due to orthonormality, the return probability
\begin{equation} \label{pr}
p_{a,x_{0}} \equiv |c_{a,x_{0}}(t)|^2 = [1+|v_{k',x_{0}}|^{2}(\cos{gt} -1)]^2. 
\end{equation}

In other words, 
the amount of information released by the initial excitation ultimately depends on the overlap between
$\ket{\beta_{k'}}$ and $\ket{e_{x_{0}}}$.
For a uniform array, which is our case, 
the free-field spectrum 
consists of plane waves of the form
$v_{k,x} \propto \sin{(kx)}$, with $k = \pi m / (N+1)$ and $m=1,\ldots, N$, and so
the overlap should be
small enough to retain most of the amplitude.
Still, for finite $N$ some amount of atomic probability periodically flows out 
of the initial 
state reaching the other atomic states in phase as
\begin{equation} \label{px}
p_{a,x} = |v_{k',x}v_{k',x_{0}}^{*}|^2(\cos{gt} -1)^2
\end{equation}
for $x\neq x_{0}$.


Taking the other limit, that is when we increase $g/J$ until reaching the strong atom-field coupling regime 
\cite{makin2009, almeida2013}, every normal mode becomes fully dressed
and the corresponding eigenstates effectively take the form
$\ket{\psi_{k}^{\pm}} =  (\ket{\alpha_{k}} \pm
\ket{\beta_{k}})/ \sqrt{2}$. These are polaritons that
form two single-particle dispersion branches having the
very same structure as of an embedded array with
the hopping scale redefined by $J/2$ [see Eq. (\ref{dressed_energy})]. 
A much better way to visualize this is by rewriting the JCH Hamiltonian, Eq. (\ref{jchhamilt}) in terms of local polaritonic operators $P_{x}^{(n,\pm)} \equiv \ket{\emptyset}_{x}\bra{n\pm}$ \cite{angelakis2007}. Dropping out terms with $n \neq 1$ the Hamiltonian becomes (recall we set $\omega_{c}=0$)
\begin{align} \label{Hpol}
H &= g\sum_{x=1}^{N}\left(P_{x}^{(+)\dagger}P_{x}^{(+)} - P_{x}^{(-)\dagger}P_{x}^{(-)}\right)\nonumber \\ 
&- \dfrac{J}{2}\sum_{x=1}^{N-1}\left(P_{x}^{(+)\dagger}P_{x+1}^{(+)}
+P_{x}^{(-)\dagger}P_{x+1}^{(-)}\right. \nonumber \\ 
&\left. +P_{x}^{(+)\dagger}P_{x+1}^{(-)}+P_{x}^{(-)\dagger}P_{x+1}^{(+)} + \mathrm{H.c.}\right),
\end{align}
where $P_{x}^{(\pm)} \equiv P_{x}^{(1,\pm)}$ for brevity. 
The above Hamiltonian is equivalent
to a double tight-binding array connected to each other through the hopping terms that exchange between even ($\ket{+}_{x}$) and odd ($\ket{-}_{x}$) polaritons [see Fig. \ref{fig1}(d)].
This can be further simplified in the strong coupling regime ($g\gg J$), where
both chains become effectively decoupled \cite{angelakis2007,makin2009},
i.e., those inter-converting terms are fast rotating and
can be dropped out. It is worth mentioning that
the even and odd polaritonic operators obey, each, the same algebra as the spin$-1/2$ ladder operators. Therefore, in this regime, the JCH Hamiltonian effectively describes a $XY$ spin chain with spin up (down) corresponding to the presence (absence) of polaritons \cite{angelakis2007,bose2007}.

In such strong atom-field coupling scenario
the dynamics of the atomic excitation 
mimics that of a single particle 
propagating along either of the uncoupled effective chains (it spreads out ballistically in a uniform chain) with hopping constant $J/2$, the only difference being that
it is continuously converted back and forth to a photonic state
at rate $g$ \cite{makin2009}.
Time-evolved atomic coefficients in this case read
\begin{equation}\label{at_amp2}
c_{a,x}(t) =\cos{gt} \sum_{k} e^{-i\frac{\omega_{k}}{2}t}v_{k,x}v_{k,x_{0}}^{*}.
\end{equation}
Also note in Fig. \ref{fig1}(d) that if the system is initialized in an even (odd) polariton state, the odd (even) 
counterpart will not take part in the dynamics.

\section{\label{sec4} Entanglement properties}

Now that we have made an overall analysis of the
two main limiting regimes of the JCH model, we 
are to track 
the entanglement over time between atomic and photonic degrees of freedom via the von Neumann entropy as well as 
between pairs of atoms
via the concurrence. Those measures are introduced next.

\subsection{Entanglement measures}

The single-excitation subspace is spanned by $\lbrace \ket{1_{i}},\ket{e_{i}} \rbrace$ so that a general state can be written as
\begin{equation}
\ket{\psi} = \sum_{i=1}^{N}\left(c_{f,i}\ket{1_{i}}+c_{a,i}\ket{e_{i}}\right),
\end{equation}
where $c_{f,x}$ and $c_{a,x}$ are the field and atomic coefficients, respectively. 
In the density-operator formalism, we have
\begin{align}
\rho = \ket{\psi}\bra{\psi} = &\sum_{i=1}^{N}\sum_{j=1}^{N}
( c_{f,i}c_{f,j}^{*}\ket{1_i}\bra{1_j}
+c_{f,i}c_{a,j}^{*}\ket{1_i}\bra{e_j} \nonumber \\ 
&+c_{a,i}c_{f,j}^{*}\ket{e_i}\bra{1_j}
+c_{a,i}c_{a,j}^{*}\ket{e_i}\bra{e_j}).
\end{align}
Now, tracing out the cavity (field) modes,
$\rho_{a} = \mathrm{Tr}_{f}[\rho]$, we obtain
\begin{equation}
\label{rhoa}
\rho_{a}
= \sum_{i=1}^{N} \vert c_{f,i}\vert^{2} \ket{\Downarrow}\bra{\Downarrow}
+\sum_{i=1}^{N}\sum_{j=1}^{N}c_{a,i}c_{a,j}^{*}{\sigma}_{i}^{+}\ket{\Downarrow}
\bra{\Downarrow}{\sigma}_{j}^{-},
\end{equation}
where $\ket{\Downarrow}\equiv\ket{g}_{1}\ldots\ket{g}_{N}$. 

Note that, in general, $\rho_{a}$ is a mixed state and thus
the atomic component, as a whole, is said to be entangled with the photonic subsystem. The diagonal form of $\rho_{a}$ 
has only two entries, $\Pi_{f} \equiv \sum_{i}\vert c_{f,i} \vert^2$ and $\Pi_{a} \equiv \sum_{i}\vert c_{a,i} \vert^2$, namely the total photonic and atomic probabilities, respectively.
Since $\ket{\psi}$ is a pure state, we can evaluate the amount of entanglement between 
two partitions through the von Neumann entropy. 
For the, say, atomic component,
\begin{equation}
   S[\rho_{a}] = -\mathrm{Tr}\rho_{a}\mathrm{log}_{2}\rho_{a} 
   = -\Pi_a\mathrm{log}_{2}\Pi_a - (1 - \Pi_a)\mathrm{log}_{2}(1 - \Pi_a),
\end{equation}
which gives 0 (1) for a fully separable (entangled) state. 
Note that the entropy reaches its maximum for $\Pi_{f}=\Pi_{a}=1/2$, that is 
$S_{\mathrm{max}} = - \mathrm{log}_{2}(1/2) = 1$.

To evaluate bipartite entanglement between the atoms we choose a 
pair of sites, say, $i$ and $j$ and further trace out the rest of them from $\rho_{a}$ 
[Eq. (\ref{rhoa})] to obtain a four dimensional reduced matrix spanned in the basis $\lbrace \ket{gg}, \ket{ge}, \ket{eg}, \ket{ee} \rbrace$,
\begin{equation} \label{rho_ij}
\rho_{i,j} = 
\begin{bmatrix} 
1 -|c_{a,i}|^{2}-|c_{a,j}|^{2} & 0 & 0 & 0\\ 
 0& |c_{a,i}|^{2} & c_{a,i}c_{a,j}^{*} & 0\\ 
 0&c_{a,j}c_{a,i}^{*}  & |c_{a,j}|^{2} & 0\\ 
 0& 0 & 0 & 0
\end{bmatrix}.
\end{equation}
%
Despite the fact that it is not straightforward to evaluate the entanglement of a mixed state, a simple
expression does exist for an arbitrary state of two qubits.
The so-called concurrence is defined by \cite{wootters1998} 
\begin{equation}
C(\rho_{i,j}) = \mathrm{max}\lbrace 0, \sqrt{\lambda_{1}}-\sqrt{\lambda_{2}}-\sqrt{\lambda_{3}}-\sqrt{\lambda_{4}} \rbrace,
\end{equation}
where $\lbrace \lambda_{i} \rbrace$ are decreasing eigenvalues, of the matrix $\rho_{i,j} \tilde{\rho}_{i,j}$, with
\begin{equation}
\tilde{\rho}_{AB} = (\sigma_{y}\otimes \sigma_{y})\rho_{AB}^{*}(\sigma_{y}\otimes \sigma_{y})
\end{equation}
and $\rho_{i,j}^{*}$ being the complex conjugate of $\rho_{i,j}$, and $\sigma_{y}$ the Pauli operator.
For a separable (fully-entangled) state $C = 0$ ($C=1$). 
Evaluating for Eq. (\ref{rho_ij}), we get
\begin{equation}\label{c_mn}
C_{i,j} \equiv C\left(\rho_{i,j}\right) = 2\vert c_{a,i}c_{a,j}^{*}\vert =2\vert \langle e_{i} \vert \psi \rangle \langle \psi \vert e_{j} \rangle \vert.
\end{equation}

\subsection{Time evolution}

The protocol starts with a single atomic excitation prepared in
the middle of the coupled-cavity system and we let it evolve 
naturally as $\ket{\psi(t)} = e^{-iHt}\ket{e_{x_{0}}}$, with $x_{0} = \frac{N+1}{2}$ and $N$ being odd 
so as to have a mode at the center of the band. We keep $\omega_{a}=0$ henceforth. 
Note that this triggers a JC-like interaction between atomic and field modes
at that level when in the weak-coupling regime,
as discussed in the previous section. 
Also note that [cf. Eq. (\ref{at_amp})] 
$v_{k,\frac{N+1}{2}} \propto \sin{\pi m/2} =0$ for even $m$.
As the resonance is set at the center of the band, $m = (N+1)/2$ must be an odd number, otherwise
there is no propagation when $g\ll J$. 

\begin{figure}
    \centering
    \includegraphics[width=0.5\textwidth]{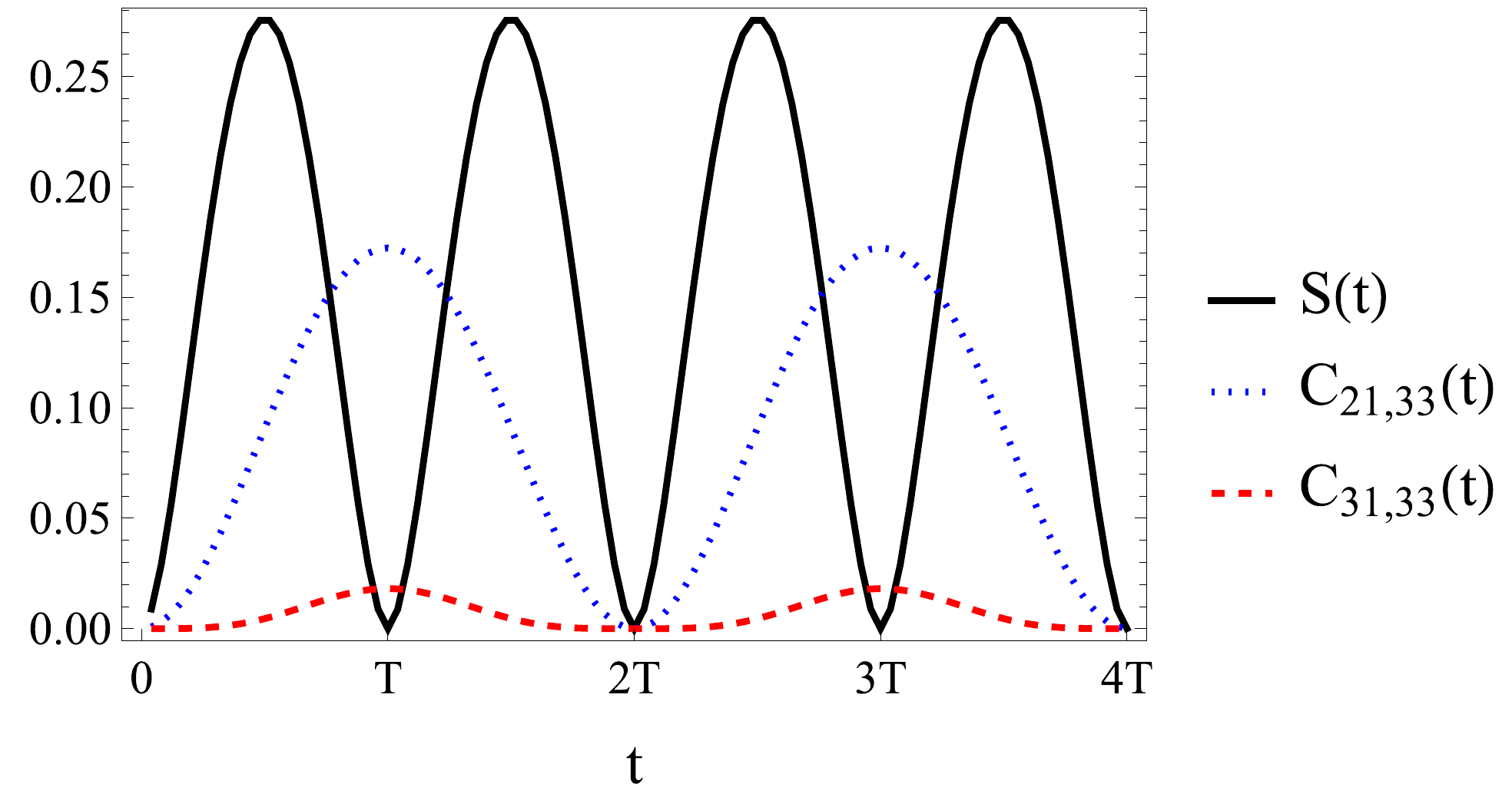}
    \caption{(Color online) Exact time evolution of the atom-field von Neumann entropy $S$ and atomic concurrences 
$C_{x_{0},33}$, $C_{31,33}$, 
for $\ket{\psi (t=0)} = \ket{e_{x_{0}}}$
with $x_{0} = 21$ on a uniform coupled-cavity array featuring $N=41$ sites operating in
the weak coupling regime with $g = 10^{-3}J$ and $\omega_{a}=\omega_{c}=0$. The entropy oscillates with period $T = \pi/g$.}
    \label{fig2}
\end{figure}

Given the fact that the atomic
wavefunction can only spread out if mediated by the field,
generation of entanglement between pairs of atoms must be preceded
by the development of atom-field correlations.
We are now to see 
how this goes for both limiting interaction regimes.
The exact entropy dynamics for the weak-coupling regime ($g\ll J$)
is depicted in Fig. \ref{fig2}
alongside concurrence for two distinct pairs of atoms. 
The total atomic probability
$\Pi_{a}(t) = 1-|v_{k',x_{0}}|^2\sin^{2}(gt)$ and thus 
the entropy evolves with period $T = \pi/g$, half that of 
the return probability in Eq. (\ref{pr}). 
So, two entropy cycles cover (from the beginning) 
the release of energy from $\ket{e_{x_{0}}}$ to the photonic degrees of freedom, 
followed by excitation of the remaining atomic states [see Eq. (\ref{px})] at $t=T$ (when $S = 0$),
ending up with full recovery of the initial state at $t=2T$ 
via a second transition through
the photonic mode.    
Atomic concurrences set along within the same timescale, reaching its maximum at times $t = T,3T,5T,\ldots$ in phase,
as already implied in Eq. (\ref{px}).
In general, it is crucial to highlight that degree of entropy generation, as well as 
the precise timing of the maximum concurrence are governed by the overlap $v_{k',x_{0}}$ since communication between atomic and photonic degrees of freedom in the weak coupling regime involves exchange between $\ket{\alpha_{k'}}$ and $\ket{\beta_{k'}}$ at a single level $k'$, rather than over the full spectrum. 
Another feature to note in Fig. \ref{fig2} -- also by a careful inspection of Eq. (\ref{at_amp}) -- 
is that the concurrence involving 
the atom located at the initial site $x_{0}$ overcomes entanglement between any other
pair (Fig. \ref{fig2} shows that for two representative pairs).   
This is due to the spectral profile of 
the uniform coupled-cavity array 
for it restricts the flow out of $\ket{e_{x_{0}}}$ thereby
leaving the remaining cavities
with limited resources to establish atomic entanglement, especially for larger $N$. 


\begin{figure}[t]
    \centering
    \includegraphics[width=0.3\textwidth]{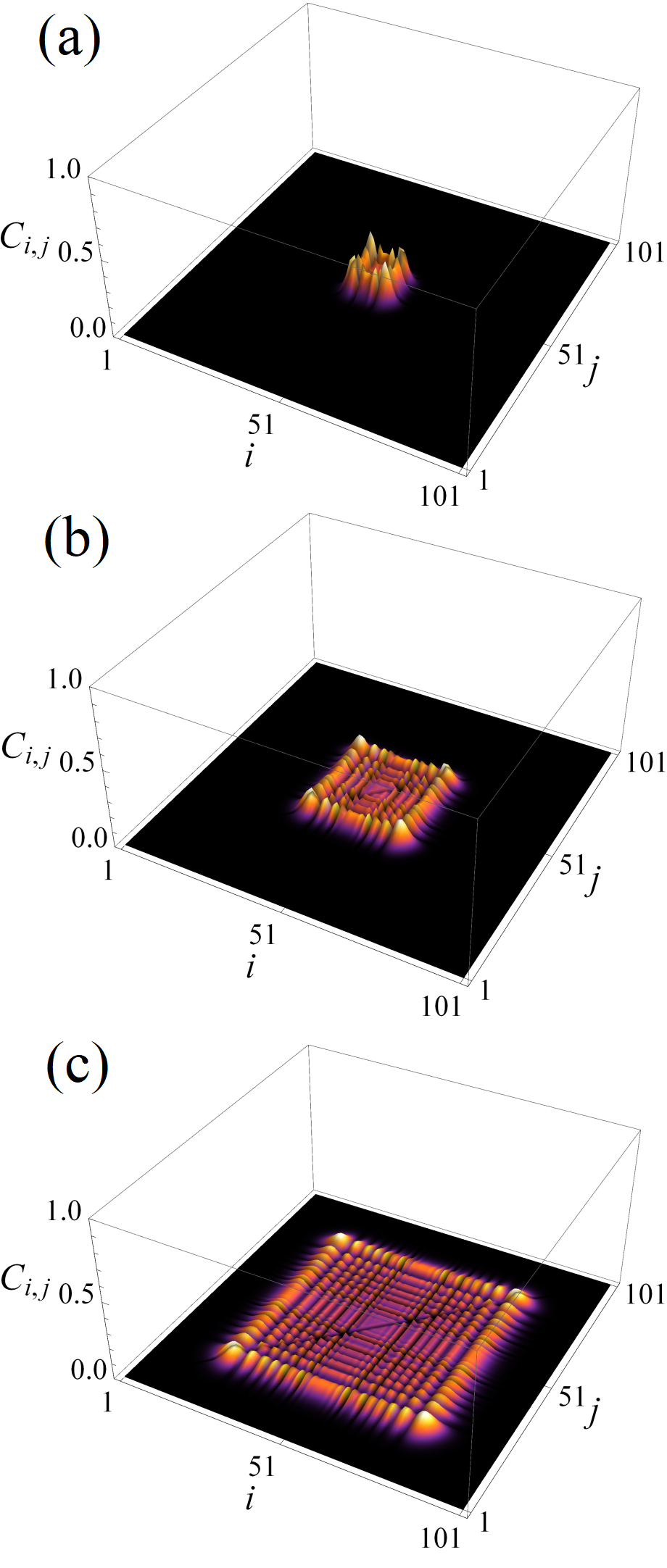}
    \caption{(Color online) Snapshots of the atomic concurrence distribution $C_{i,j}(t)$
in the strong coupling regime for (a) $t = 2000 \pi/g$,
(a) $t = 5000 \pi/g$, and (c) $t = 10000 \pi/g$, with $g = 10^3 J$, such that $\Pi_{a}(t) = \cos^{2}(gt) =1$ (thus $S(t) = 0$).
The system consists of $N=101$ coupled cavities with the initial state
being $\ket{\psi (t=0)} = \ket{e_{51}}$ and $\omega_{a} = \omega_{c} = 0$ and results 
are exact as obtained directly from Hamiltonian (\ref{jchhamilt}). Note that the atomic wavefunction propagates at rate $J/2$ and thus the front pulse
roughly advances a site per $J^{-1}$ elapsed time.}
    \label{fig3}
\end{figure}

Moving on to the the strong coupling regime ($g\gg J$), we get a whole different picture. 
Now, there is no special mode triggering the dynamics. All the modes are involved and atomic 
degrees of freedom are completely mixed with their photonic analogs. 
Assisted by photonic scattering, 
the initial atomic excitation 
spreads out ballistically at rate
$J/2$, as $\ket{e_{x_{0}}}$ is a superposition 
of even and odd polaritons, both spanning
the uncoupled effective chains seen in Fig. \ref{fig1}(d).
As it propagates, the atomic wavefunction is constantly mirrored back and forth to its photonic form in a much faster timescale. 
In this limit, the entropy is fed with total atomic probability $\Pi_{a}(t) = \cos^{2}(gt)$, implying
that $S(t)$ reaches maximum at times $t = m \pi/(4g)$ for odd $m$ (that is when $\Pi_{a}(t)=1/2$).
Note that the above property is general in that it holds for any size $N$ and hopping scheme, with the resulting atomic
dynamics always obeying the underlying spectral properties of the coupled-cavity array, as long as $g$ is larger than the
free-field band width as well as $\omega_{a}$.  
Therefore, given that entropy generation is local, generation of atomic entanglement is ultimately driven
by wave dispersion.
Figure \ref{fig3} shows some snapshots of the concurrence distribution 
at times when $\Pi_{a}(t)=1$ to get the most of $C_{i,j}$. As one should expect, entanglement is well distributed throughout the array as it evolves, with
stronger correlations taking place within each front pulse as well as between them.

\begin{figure}
    \centering
    \includegraphics[width=0.5\textwidth]{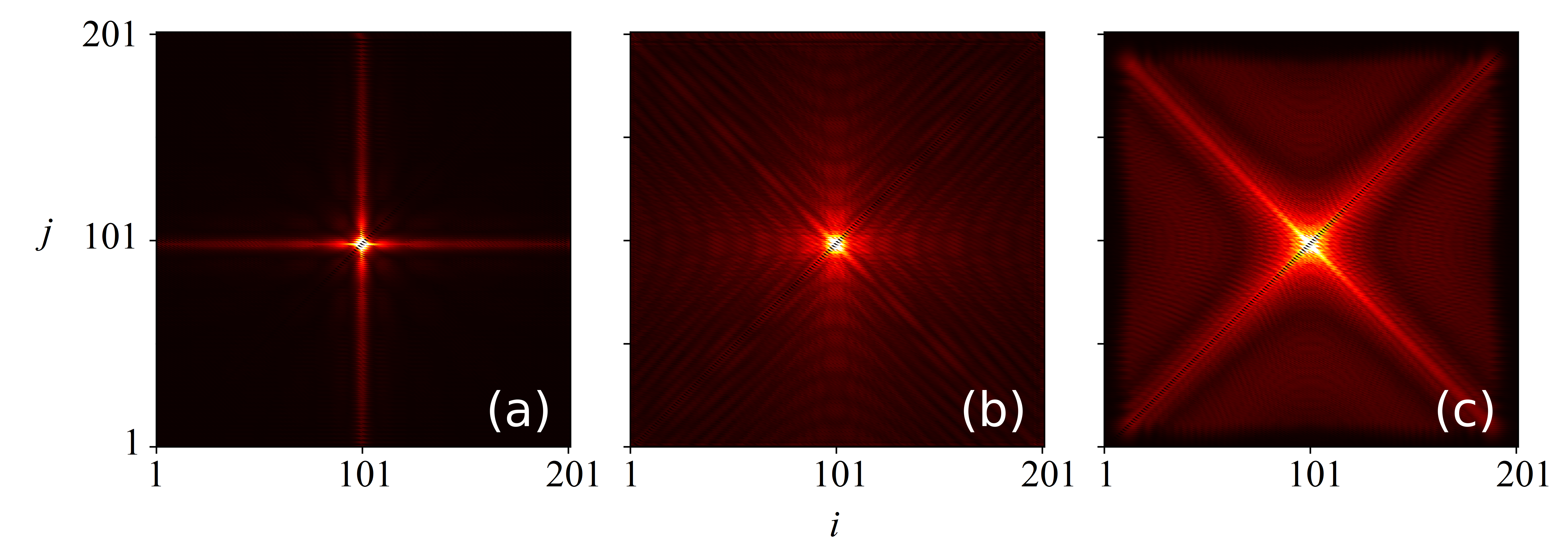}
    \caption{(Color online) Maximum concurrence $C_{i,j}(t)$ between all pairs of atoms recorded within 
    time interval $tJ \in [0,90]$ for three distinct regimes represented by (a) $g = 0.1J$, (b) $g=1.5J$, and (c) $g = 10J$,
    considering $N=201$ cavities and $\ket{\psi(t=0)}=\ket{e_{101}}$. 
    The time window was chosen so that the wavefunction did not reach the boundaries in (c).
   Color map goes from dark to bright as $C_{i,j}(t) \in [0,0.25)$. 
    }
    \label{fig4}
\end{figure}

Finally, to get a better glance over the spatial distribution of atomic entanglement, in Fig. \ref{fig4} we display
the maximum concurrence recorded within a fixed time interval 
for all $C_{i,j}$ ($i\neq j$) and
covering three different regimes.
In the weak coupling scenario, as a single atomic excitation prepared above the ground state
of a uniform coupled-cavity array undergoes a trapping mechanism \cite{makin2009, ciccarello2011, almeida2016},
it pairs up with each of the remaining atoms to produce  
the entanglement pattern we see in Fig. \ref{fig4}(a). In this situation, we shall remember that entanglement
does \textit{not} spread out from the center of the array [as in Fig. \ref{fig3}]; it is generated all at once
as the entropy dynamics involves resonant interaction between atomic and photonic delocalized modes (cf. Fig. \ref{fig2}).    
We observe that such spatial pattern similar to that of disordered chains reported in Ref. \cite{almeida2017}, what is very 
interesting as our array is fully uniform. It means that the atomic trapping mechanism can be thought as a sort of interaction-induced localization. 

Setting up a moderate interaction strength ($g \sim J$), the entanglement distribution in Fig. \ref{fig4}(b) 
does not seem to display a very definite pattern but it 
marks a crossover to the strong coupling regime shown in Fig. \ref{fig4}(c).
This one is highlighted by the onset of stronger correlations between neighboring atoms as well as 
between atoms equidistant from the center of the array,
as already suggested by Fig. \ref{fig3}. 
As a full band of extended states begins to take over the dynamics as $g$ is increased, 
the initial atomic excitation rapidly communicates with the photonic degrees of freedom locally 
and spreads out simulating the dynamics of single photon in a atom-free coupled-cavity array with $J$ 
replaced by $J/2$ in the limit $g \gg J$. Those maxima in Fig. \ref{fig4}(c) are thus 
recorded when the front pulse of the atomic wavefunction passes by \cite{almeida2017}.  
%

\section{\label{sec5}Concluding remarks}
We have studied entanglement generation and 
its spatial distribution control over a 1D uniform coupled-cavity array described by
the JCH Hamiltonian in the single-excitation sector. 
We carried out detailed analytical calculations for two limiting regimes so as to gain intuition
over how photonic and atomic degrees of freedom are combined when put to interact via the JC Hamiltonian. 

With all that set, we moved on to study entanglement generation via time evolution of a single atomic excitation
prepared above the vacuum (ground) state. 
We focused on
the von Neumann entropy between atomic and field states and the concurrence between pairs of atoms. 
We found that, in the weak coupling regime ($g \ll J$), 
entropy generation follows the same timescale as that of concurrence and depends directly on the likelihood 
of energy release from the emitter located at the initial cavity -- which, in turn, depends on the resonant field mode --, 
thus being crucial to make resources available to the other atoms to build up correlations.
Because of the atomic trapping behavior, entanglement between
the initial atom and the remaining ones prevails over that involving any other pair.
This sort of interaction-induced localization occurring in
the weak coupling regime is certainly worth to be further investigated in other
scenarios, such as beyond the single-excitation subspace where the photon-blockade sets in. 
Things should also become more involved when considering, for instance, complex networks \cite{almeida2013}
and other lattices with unique topological features \cite{ciccarello2011, almeida2016}, where localized modes are built-in. 
 
In the strong coupling regime ($g\gg J$), entanglement dynamics is more straightforward
as the entropy oscillates (now between minimum and maximum) much faster than the actual propagation of the atomic wavefunction, meaning
that entropy generation is strictly due to local interactions, differently from the weak coupling limit.
Atomic concurrence then builds up depending on the dispersion profile of the embedded 
array at rate $J/2$. A uniform one entails ballistic spreading and so the amplitudes are concentrated 
within the front pulse. Higher degrees of pairwise entanglement are then to be found 
in between nearest-neighbor atoms and between them and their equidistant counterpart at the other side
of the array in respect to its center. 

Although we found that long-distance atomic entanglement 
becomes weaker due to dispersive effects of the array itself, it may be distilled into pure singlets \cite{horodecki98}, 
to be used in, e.g., quantum teleportation protocols.
The natural dynamics of the JCH Hamiltonian may thus be 
harnessed to generate entanglement between distant nodes
in hybrid light-matter quantum network architectures \cite{kimble2008, ritter2012}.


%

\end{document}